\documentclass[%
 reprint,
 amsmath,amssymb,
 aps,
floatfix,
]{revtex4-1}

\usepackage{graphicx}
\usepackage{dcolumn}
\usepackage{bm}
\usepackage[english]{babel}

\usepackage[utf8x]{inputenc}

\usepackage{geometry}		
\usepackage{graphicx}
\usepackage[above,below]{placeins}	
\usepackage{times}
\usepackage{float}
\usepackage{natbib}
\usepackage[bottom]{footmisc}

\begin{document}

\title{Long-term collaboration with strong friendship ties improves academic performance in remote and hybrid teaching modalities in high school physics}

\author{
Javier Pulgar$^{1,2,*}$, Diego Ram\'irez$^{2}$, Abigail Umanzor $^{2}$, Cristian Candia$^{3,4}$, Iv\'an S\'anchez$^{1,2}$\\
\small{$^{1}$Departamento de F\'isica, Universidad del B\'io B\'io, Concepci\'on, 4051381, Chile.}\\
\small{$^{2}$Grupo de Investigaci\'{o}n en Did\'{a}ctica de las Ciencias y Matem\'{a}tica (GIDICMA)}\\
\small{$^{3}$Data Science Institute, Facultad de Ingeniería, Universidad del Desarrollo, Las Condes, 7610658, Chile.}\\
\small{$^{4}$Northwestern Institute on Complex Systems (NICO), Northwestern University, Evanston, IL 60208.}\\
\small{$^{*}$Corresponding Author: jpulgar@ubiobio.cl}
}
\date{\small{\today}}








\begin{abstract}
Collaboration among students is fundamental for knowledge building and competency development. Nonetheless, the effectiveness of student collaboration depends on the extent that these interactions take place under conditions that favor commitment, trust, and decision-making among those who interact. The sanitary situation and the transition to emergency remote teaching have added new challenges for collaboration, mainly because now, students' interactions are wholly mediated by Information and Communication Technologies. This study first explores the effectiveness of different collaborative relationships on physics from a sample of secondary students from two schools located in rural and urban areas in southern Chile. We used Social Network Analysis to map students' academic prestige, collaboration, and friendship ties. We define a strong association if two students are declared mutually as friends. Using multiple linear regression models, we found positive effects of collaboration on grades, particularly among students working with friends (strong ties). Then, the second study corresponds to a quasi-experiment on four classes at the analyzed urban school during the following academic year. Here, students attended hybrid sessions to test whether the effects of collaboration found on the first study were stable throughout two semesters. We followed the same procedures as in the exploratory phase. We found that the initial effects of collaboration on grades are negative for all collaborative variables, while strong working ties afford academic gains later into the year. These results contribute to the literature of collaboration in physics education and its effectiveness, taking into account social relationships and the needed time for developing beneficial collective processes among students in the classroom. We discuss these results and their implications for instructional design and guidelines for constructive group-level processes.

\end{abstract}

\maketitle

\section{Introduction}
Student collaboration is increasingly gaining attention for its benefits on learning and overall human development \cite{echeita2012, Barkley_ColTech, cerda2019}. Education scholars and policy makers have highlighted teamwork and social competencies as key abilities for life and work in the XXI Century \cite{Bao2019, pllegrino}. Using collaborative learning practices in education harnesses students' understanding of science and technology, analytical skills and openness to diversity \cite{cabrera_cl_2002}. This emphasis on social skills is necessary to equip individuals with skills for learning, adaptation, communication and creativity in the face of complex and multidisciplinary challenges  \cite{sawyer_edu_innovation}. The worldwide pandemic due to COVID-19, has without a doubt added external difficulties to the development of collaborative skills, as schools and universities were forced to move from face-to-face to remote instruction, and with the subsequent changes in students' roles and means for socialization and interaction. 

Remote teaching implies that students' communication is mediated via Information Communication Technologies (ICTs) \cite{Vonderwell2005,Traxler2018, panigrahi}, such as forums, chats, emails, or video conferences, besides informal channels of communication such as instant message applications and related technologies. Even though technologies are ubiquitous in today's society, the effectiveness of remote education strongly depends on their accessibility to technology, and on their readiness to navigate and deal with these with autonomy and self-regulation \cite{Hung2010, Kebritchi2017}. These conditions clearly affect how students' engage in forums or chats, instances that become proxies of social engagement and collaboration  \cite{Romiszowski2004, Wise2013, Traxler2018}, as individuals share/access information posted by their peers. 

The effectiveness of social interactions for learning and academic performance are linked with having appropriate collaborative skills and finding the right partners to work with\cite{le_collaborative_learning_practices}. Identifying effective working ties implies assessing a myriad of variables related to the nature of the learning activity and its requirements \cite{pulgar_2021}, along with the trust and commitment embedded in one's working relationships \cite{Pulgar_TSC}. Here, the use of Social Network Analysis (SNA) affords relevant methodological and theoretical tools to capture the different types of social and collaborative relationships observed in the classroom, to later test whether these foster (or hinder) students' performance consistently throughout the year.  

With the goal to extend our understanding of effective collaboration in physics classrooms in the midst of extraordinary academic years due to COVID-19, we explore the dynamic academic effects of having working ties with friends and/or other prestigious students. In detail, on a first exploratory phase we tested the academic effects of different collaborative ties (i.e., strong, weak and with good students) at the end of the academic year during an online teaching modality in high school physics. Secondly, we conducted a quasi-experiment to test whether collaboration effects were stable throughout three group activities conducted during the academic year, on a hybrid teaching modality in high school physics. On both experiments, we used students' social networks to characterized different collaborative relationships depending on whether these occurred among friends (i.e., strong), in the absence of friendship (i.e., weak), and between peers perceived as proficient in physics (i.e., academic prestige). This work contributes first, with a novel method to categorize students' working relationships from SNA, and second, with empirical evidence that shows the academic gains of various collaborative ties under distinctive teaching conditions, as well as its timely evolution during a year.

\section{Theoretical Perspectives}

\subsection{Student Networks and Performance}

Using SNA on education has enabled researchers in-depth analysis to understand the benefits of student collaboration from the logic of social capital, in the sense the number and strategic social interactions enables access to various forms of resources and information, and the potential to adopt behaviors related to success \cite{Grunspan, Putnik}. Good performance has been found associated with students' social networks on face-to-face \cite{Bruun2013, Pulgar19, Putnik, Brewe_Sawtelle2012} and remote courses \cite{Morris2005, Traxler2018,Dawson2008}, while covering a wide range of topics and strategies. Such findings align with the sociocultural view of cognitive and human development, as knowledge is constructed while learners work alongside others who could potentially expand the frontiers of their individual achievements, on what has been defined as the Zone of Proximal Development (ZDP) \cite{vyg}. 

The benefits of having multiple interactions with peers on the classroom, and becoming a central member of the network extends to other variables related to achievement, such as retention \cite{Williams,Zwolak_Zwolak, Zwolak}, and self-efficacy, or the believe on one's abilities to successfully meet the academic expectations \cite{Dou2016}. Besides, programs grounded on the principles of collaborative learning have shown important results for developing social skills \cite{Carrasco2018}, and trust among team members, which increases the likelihood of group effectiveness \cite{leon2017}. 

Nonetheless, more recent evidence on SNA on university physics courses suggests that the academic gains from having multiple working ties depend on the nature of the learning task \cite{Pulgar19}. Accordingly, higher degree centrality is associated with a worse performance on well-defined physics problems (e.g., textbook problems), whereas open-ended and creative activities benefit from multiple social ties outside one’s group (i.e., brokering knowledge \cite{Burt2004,Pulgar19}). Additional results on primary education found that the reciprocity of students' relationships becomes the necessary condition for higher achievement \cite{Candia}. This evidence supports the claim that the academic gains afforded by social interactions are sensitive to both contextual and the personality of its constituents.  

Furthermore, according to evidence on sociology and networks, the nature of the social relationship plays a key role in acquiring and developing knowledge depending on its complexity \cite{Granovetter, hansen}. For instance, new tacit, factual or \textit{knows what} type of information is easily accessible through weak social ties, that is, among individuals who do not share a sense of commitment nor are deeply embedded into the relationship \cite{Granovetter}. The key consideration is that weak ties are observed between individuals who do not belong to one's cohesive group -governed by strong relationships-, and are therefore preferable for accessing simple new ideas outside such networks. On the contrary, strong ties are better suited for learning complex and non-tacit ideas, because developing such knowledge requires a social commitment and a common language to transfer the intricacies of this new information \cite{hansen}. Such learning considerations are frequently observed among individuals who share a strong relationship, like friendship. Plus, informal friendship ties are better at disseminating behaviors linked with high performance \cite{dokuka}. Yet, evidence on the challenges for effective collaboration have found that working with friends hinders individual responsibility and the seriousness in the construction of arguments \cite{le_collaborative_learning_practices}, a finding that mirrors the negative effects of high network cohesion \cite{wise}, which might lead to underestimating the complexity of the task through high team efficacy \cite{park}. 

All of the above stresses out the challenges of guiding students' collaboration in the classroom, taking into account the nature of the tasks and the information to be learned. These complexities are now relevant in this transition to remote teaching, where students' interactions are mediated by ICTs \cite{Traxler2018}. Studies on remote learning have conceptualized students' course participation using traditional methods, such as the number of posts on online forums \cite{Romiszowski2004, Vonderwell2005}, the time dedicated to reading comments \cite{Hrastinski2009,Wise2013}, or via content analysis of the posts written \cite{deweber}. Recently, researchers have used SNA to explore different forms of participation on online courses. For instance, Traxler, Gavrin and Linden \cite{Traxler2018} conceptualized students' co-occurrence of posts on discussion forums, which were defined as the links on the participation network, and later used to identify students' centrality as a proxy in the access to the content-related ideas. Under this condition, higher access to information on forums, that is, high network centrality is positively correlated with academic success \cite{Traxler2018}. Through similar network methods, an additional study has found positive correlations between students' participation on online courses -centrality- with their sense of belonging \cite{Dawson2008}. Yet, the centrality on this participation network offers positive effects specially to high performance students, as content analysis has shown that their discussions circled around conceptual aspects of the course content, whereas the discussion networks formed by low performing students tend to focus on practical and superficial elements of the curriculum \cite{dawson2010}. Although these experiences have sought to map underlying structures in the access to information through the co-occurrence of posts, for instance, in the studies we will describe later, we instead pay attention to students' declared social relationships. In the next section we dive into the interplay between teaching strategies and forms of collaboration found in relevant literature.

\subsection{Teaching Strategies and Students Collaboration}


From the perspective of learning methodologies, active classrooms have shown higher levels of social interactions compared to traditional lecture-based instruction \cite{traxler2020,Commeford2020,Pulgar19, Pulgar_Sochifi,Brewe_Kramer2012}. Student-centered classrooms favor autonomy over the learning process through peer interaction, because these tend to include activities designed to encourage decision-making, content manipulation and communication, thus allowing students to perceive learning as a construction mediated by collaboration \cite{pulgar_2021, Pulgar19}. The social and cognitive attributes of activities implemented on active learning methodologies are also recommended on virtual classrooms \cite{chametzky, Luyt2013, Niess2013, gupta2013}, because these encourage higher levels of academic achievement and digital competencies \cite{merono2021}. 

The work of Johnson, Johnson and Holubec \cite{Johnson_Johnson} on student collaboration adds fundamental conditions for group effectiveness: positive interdependence, or the belief that the success is a collective rather than individual effort; and personal responsibility for learning and engaging on one's tasks. According to social network research on education, the above conditions are modulated by the nature of the learning tasks. For instance, well-structured physics problems hinders positive independence, as students report addressing these activities by engaging in social interactions aimed at finding the right equation or variable to use \cite{pulgar_2021}. On the contrary, real-world problems \cite{Fortus} such as generative activities \cite{pulgar_2021}, motivate collaborative strategies that align with the mentioned characteristics for effectiveness. Similar findings point out to different levels of social engagements depending on students' perceived academic status. Here, in the face of creative activities, low achieving students tend to be more active in reaching out to their high performing peers \cite{elementary_network_2017}. The interplay of social ties and learning activities is also observed in lab practices, where students displayed limited number of interactions between groups in traditional labs compared to reformed labs \cite{lab_collaboration2022}, which is associated with brokering knowledge and a key process for creative solutions \cite{Burt2004}.   

The embedded attributes of the learning activity and the socially constructed norms in the classroom are some of the reasons why one could witness various forms of collaboration between different types of activities. First, Well-structured and traditional learning activities in physics and other disciplines tend to be perceived as non-additive tasks \cite{Steiner1966}, given that when worked on groups these are likely solved by the most capable team member (i.e., academically prestigious students). These activities also relate to rather simplistic cognitive processes (e.g., application) according to a taxonomy of physics problems developed by Teodorescu and colleagues \cite{teo}. Differently, unstructured tasks, such as generative problems \cite{pulgar_2021} are intrinsically additive \cite{Steiner1966}, because they require collective efforts for content manipulation, decision-making and manipulation, and thus demand high-level cognitive skills \cite{Anderson2001, teo}. These demands require more complex forms of interactions than simple asking for information to the most capable in the classroom. Secondly, students' collaborative ties are sensitive to socially constructed expectations for learning achievements and/or performance (e.g., good grades) \cite{vignery2020achievement,stadtfeld2019integration,rizzuto2009s}. Here, instruction that emphasizes learning achievements fosters the adoption of study habits and constructive collaboration, whereas the collective pursuit of good grades -not necessarily learning- goes along with competitive comparison and self-improvement \cite{status_vs_popularity}. Consequently, both the learning activity and the classroom climate play critical roles in encouraging either constructive or rather superficial forms of social interactions between friends and/or less known peers.

\section{Research Questions}
Considering the previous review, we are far from fully comprehending the complexity of students' interactions in the classroom, whether their effectiveness in terms of achievement is sensitive to teaching methodologies, organizational decisions in the classroom and working time. Plus, the academic effects of different types of collaborative relationships has yet to be explored in physics education research (PER), particularly from a longitudinal perspective, and with the consideration of social and academic hierarchies. In sum, two studies were conducted to answer the following research questions: 
\begin{enumerate}
    \item  (RQ1) Are there differences in the academic gains of collaborating with friends and those who enjoy academic prestige, during remote teaching in high school physics? (Exploratory experiment)
    \item (RQ2) Are the academic gains of collaborating with friends and those who enjoy academic prestige stable in time during an hybrid teaching modality in high school physics? (Quasi-experiment)
    \item (RQ3) Are there differences in the academic gains of different types of collaborative ties for student groups based on friendship or randomly assigned during an hybrid teaching modality in high school physics? (Quasi-experiment)

\end{enumerate}


\section{Methods}

\subsection{Research Context on Exploratory Analysis}
The exploratory study was conducted to answer RQ1 during the second semester of 2020 on a sample of secondary students from parallel physics classes (A \& B), and from two high schools in southern Chile (Sch-1 \& Sch-2). Table \ref{Tabla1} summarizes the schools' characteristics, showing information on the student population, number of research participants, location and education. We make explicit reference to Location (Rural or Urban) and Education (Scientific, Humanistic and Technical), as it is important to clarify the possible social and cultural variables that mediate students' access to internet and ICTs, which in the context of remote teaching are the formal means through which individuals connect. Besides, in Chile's educational system, Technical schools dedicate a great percentage of its curriculum to develop technical skills, for instance as an electrician or mechanic, and where students graduate as technicians. Differently, Scientific and Humanistic schools guide students to pursue university degree programs in diverse disciplines, such as science, engineering, social science and humanities. Research participants in both schools gave consent via an online survey designed for this study, where we explain the analytical procedures along with privacy considerations. A total of 101 students agreed to be involved in the study, with 45\% of female participants.  

\begin{table}[!t]
\begin{center}
  \caption{School attributes and characteristics: Exploratory Analysis (2020) \label{Tabla1}}
    \centering
 \begin{tabular}{p{0.9cm}cc p{1.05cm} p{3.15cm}}
\hline\hline
School       & Students & Participants & Location & Education\\
\hline 
Sch1 & \centering 650   & \centering 55 & Rural    & Scientific, Humanistic \& Technical\\
Sch2 & \centering 906   & \centering 46 & Urban & Scientific \& Humanistic \\
\hline
\end{tabular}
\end{center}
\end{table}

During 2020 and due to the COVID-19 pandemic, both schools adapted their teaching methodologies from face-to-face to online sessions introducing various ICTs. Initially on School 1, teachers began the academic year with asynchronous sessions using content videos shared via text messaging apps (e.g., WhatsApp). After two months, School 1 implemented synchronous 40 min sessions organized on MEET and CLASSROOM learning management systems (LMS). This scenario differed from School 2, where remote teaching began with 1-2 hours synchronous course sessions a week after the school canceled face-to-face instruction. Similar to School 1, these sessions were organized on the same LMS. 

In addition to the initial organizational differences, both schools organized and taught physics content following opposite learning methodologies. In School 1, the main methodology was lecture-based instruction, followed by problem solving sessions where students faced well-structured activities (e.g., textbook problems), which were later assessed on their individual work. In School 2 however, physics taught via active learning methodologies, including generative activities \cite{pulgar_2021}, where students, for instance, working on groups designed videos centered around physics concepts and phenomena. Here, there was no particular mechanism for group formation, so it is assumed students teamed up with friends and/or individual affinities. The physics classes met once a week for 40 min and 1 hour respectively for Schools 1 and 2. 

\subsection{Data Collection}
Because this exploratory study aimed to determine the academic effects of different collaborative relationships on school physics, we collected final grades from the year prior to the study (2019), along with students' grades on the second semester of 2020. In 2020, students' performance on School 1 accounted for in-class activities (e.g., problem solving) and individual testing. Physics grades on School 2 however, consisted on the assessment of the generative activities (e.g., content videos and physics posters), which includes grading their structure and organization, language and theory, and a peer evaluation of the groups' performance. 

Finally, to measure students' collaboration we use an online network survey, a reliable mechanism for network mapping \cite{Borgatti2013}. This survey was designed to gather different social relationships: friendship ties; prestige on physics; and collaboration on physics. This instrument was constructed using the class' rosters, a technique utilized in previous studies to facilitate individuals' responses \cite{Brewe_Bruun16, traxler2020, Pulgar19}.

\subsection{Data Analysis}
\begin{figure*}[!htbp]
\centering
  \includegraphics[width=0.5\textwidth]{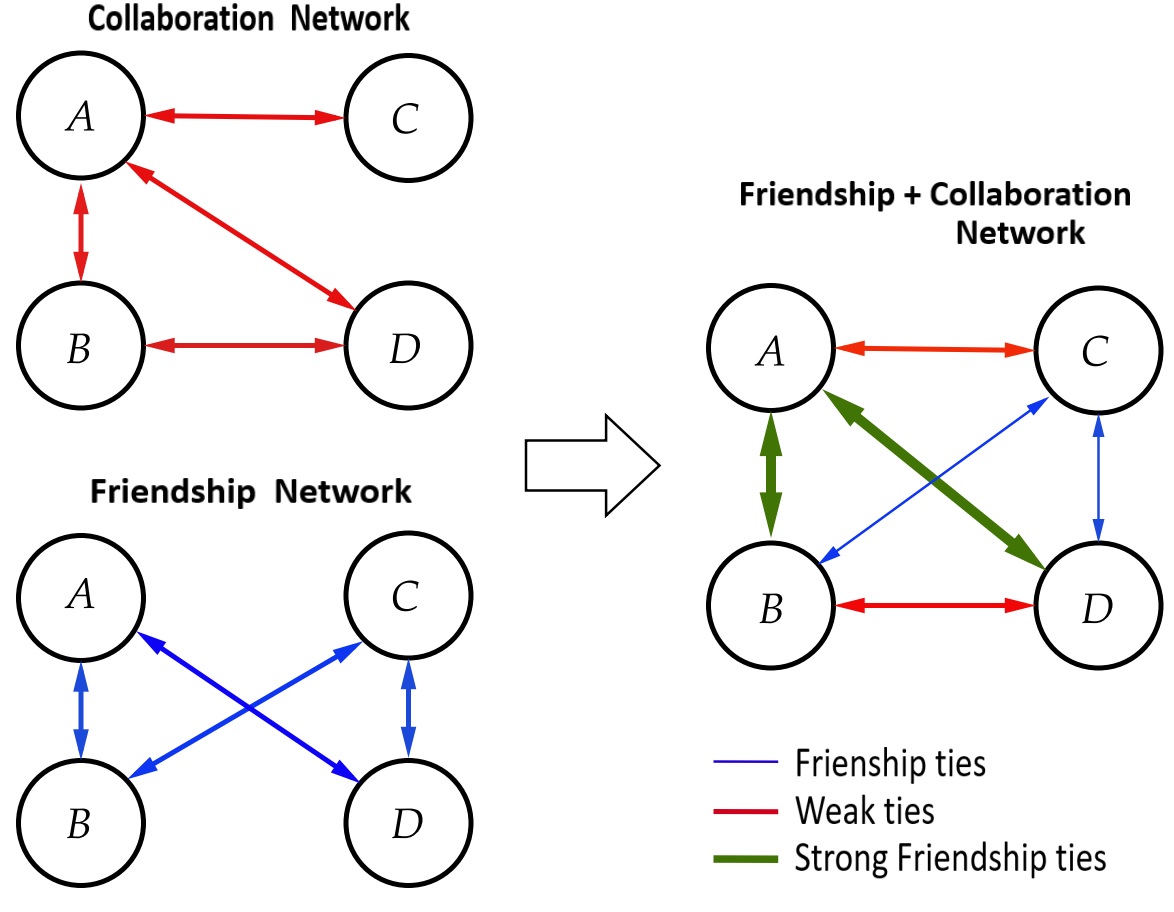}
  \caption{Diagram shows the methodology utilized to identified and construct collaborative variables number 2, 3, 4 and 5 described in Table \ref{Tabla3}. \label{methodology}}
\end{figure*}

From the network survey we identified key variables to characterize the groups in terms of their friendship relations, prestige and collaboration on both physics and mathematics courses. The survey yields indirect networks, that is, where ties are not necessarily reciprocal (e.g., A selects B as a friend, but B does not select A as a friend). Given that our analysis includes a combination of three networks and measures of degree centrality which do not include the direction of the relationship (See fig. \ref{methodology} and Table \ref{Tabla3}), we decided to account for all participants who at least had a tie in one of the three networks. Table \ref{Tabla2} summarizes the mean of key descriptive network variables for each of the social dimensions measured on the survey. The variables include the number of students per group, or nodes, the total number of ties observed, the average number of ties or average degree, and network density --percentage of observed ties considering 100\% as if all nodes on the network were connected to each other \cite{Borgatti2013,carolan}.

\begin{table*}[t!]
\begin{center}
  \caption{Descriptive statistics of directed networks: Friendship, Physics Prestige, and Collaboration on Physics.\label{Tabla2}}
  \centering
 \begin{tabular}{lcccccccccc}
\hline\hline
           &          & \multicolumn{3}{c}{Friendship}   & \multicolumn{3}{c}{Physics Prestige} & \multicolumn{3}{c}{Collaboration on Physics} \\
Group     & Nodes & Ties & Degree$^*$ & Density        & Ties & Degree$^*$ & Density              & Ties & Degree$^*$ & Density\\
\hline 
Sch1-A &  29   &  132 & 4.55 & 0.16 & 54&  1.86& 0.07&  43& 1.48&  0.05\\
Sch1-B &  26   &  95 &  3.65 & 0.15 & 67&  2.58& 0.1&  49& 1.89&  0.08\\
Sch2-A &  23   &  194 & 8.44&  0.38 & 126& 5.48 & 0.25&  52& 2.26& 0.1 \\ 
Sch2-B &  23   &  151 & 6.57  & 0.3 & 128&  5.67& 0.25&  55&  2.39&  0.1\\
\hline
\multicolumn{11}{l}{$^*$: Average number of incoming plus outgoing ties per student in the class.} 
\end{tabular}
\end{center}
\end{table*}

We operationalized collaboration as degree centrality on the collaboration network, that is, their total number of ties. Degree centrality accounts for both incoming and outgoing ties, named indegree and outdegree centrality respectively. The former indicates the number of ties headed from actors on the networks towards a focal node, whereas the latter accounts for relationships declared by the focal actor towards other nodes on the network. Besides this definition of the variable \textit{Collaboration} (see Table \ref{Tabla3}), we combined friendship, prestige and collaboration networks to identify different types of collaborative relationships among students. We turned to the strength of ties terminology used on the social network literature to differentiate between strong and weak collaborative ties \cite{Granovetter, hansen}, based on whether the observed collaborative ties occurred among students who are or not friends (i.e., strong and weak respectively). The process followed to construct these variables is depicted in Figure \ref{methodology}, and shows how we combined two networks (e.g., collaboration and friendship) to define a third network of interactions but with weighted ties, given the added friendship attribute. In the diagram, these weights are depicted in arrows of different colors and width, and represent either collaboration between friends (Green $=$ strong ties), between students who are not friends (Red $=$ weak ties), or simple friendship ties (Blue). Because our interest is placed on different types of collaboration, we counted the number -degree centrality- of these collaborative ties and saved them as individual attributes (e.g., node A has 4 strong and 2 weak ties, considering the incoming plus the outgoing ones). For the analysis of strong and weak ties we discarded simple friendship relationships. Table \ref{Tabla3} describes the five collaboration variables constructed and used for the analysis. 

\begin{table*}[t!]
  \caption{Types of collaborative relationships and definition \label{Tabla3}}
  \centering
 \begin{tabular}{p{4cm} p{12cm}}
\hline\hline
Type of Collaboration                           & Definition \\
\hline

1. Collaboration                       & Degree centrality of the collaboration network, or the total number of collaborative ties.\\

2. Strong	     & Number of collaborative ties between students who are friends. Friendship is observed on the network when at least one of the actors on the dyad indicates the other as a friend (e.g., A → B). \\

3. Weak     & Number of collaborative ties between students who do not declare themselves as friends.\\

4. Strong w/ Academic Prestige    & Number of collaborative ties between friends and who are perceived by the other as a good student on the course. Friendship is observed on the network when at least one of the actors on the dyad indicates the other as a friend (e.g., A → B). Similarly, prestige comes from its respective network and when at least one of the actors on the dyad selects the other as a good student on the course (ej., B → A, then B enjoys academic prestige).\\

5. Weak w/ Academic Prestige	& Number of collaborative ties between students who do not declare friendship ties among themselves, yet one of the actors on the dyad declares the other as a good student.  \\

\hline
\end{tabular} %
\end{table*}
We then used ordinary least squares multiple regressions models (OLS) to predict grades on both courses, while using collaboration variables as the main predictors (See Table \ref{Tabla3}). In order to isolate the effects of collaboration over grades, we include control variables like performance on the year prior to the study, School to account for variability within institutions, and gender. We also tested interactions between schools and the key predictors, and found no significant difference between schools. Finally, as a robust check, we fitted hierarchical linear models to account for the class and schools' variance. Results here did not differ from the multiple regression models.

\section{Results: Exploratory Analysis}

\subsection{Collaboration and Grades on Physics}

\begin{figure*}[!htbp]
\centering
  \includegraphics[width=0.8\textwidth]{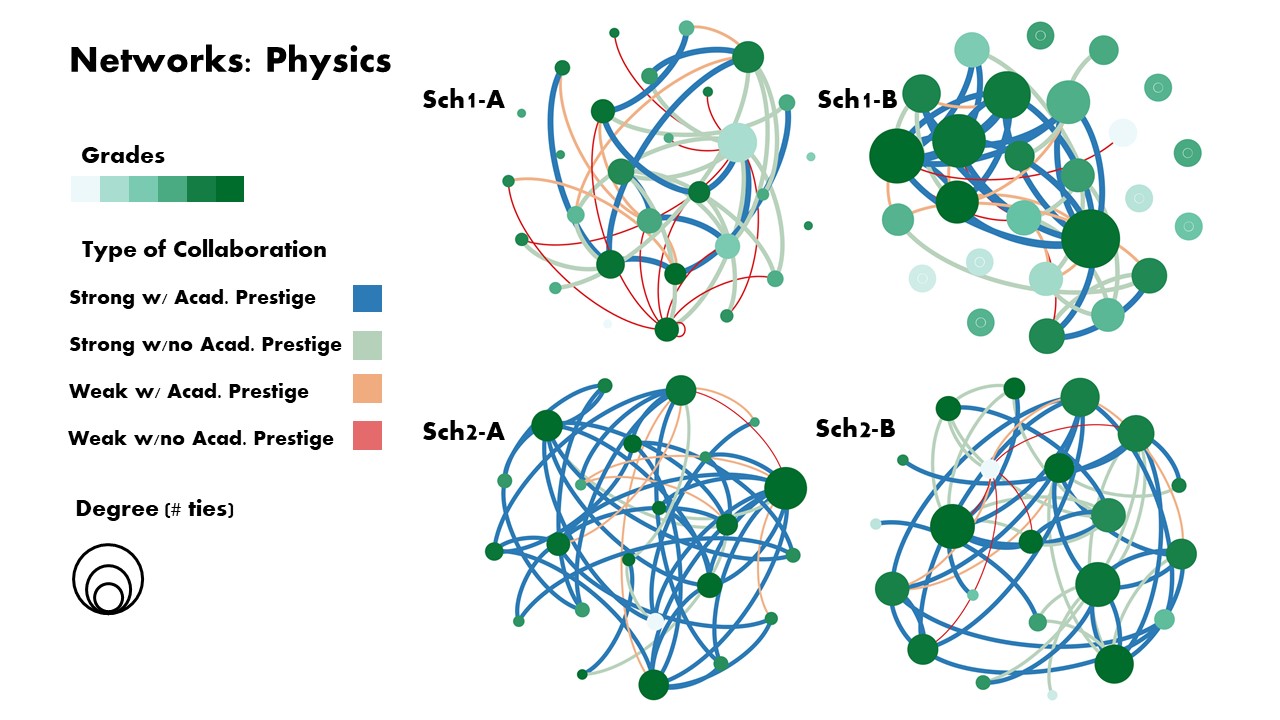}
  \caption{Collaboration network for physics. Node color represents academic performance, while its size is degree centrality --number of collaborative ties. Edge  or link colors depict the type of collaborative relationship.\label{physics_networks}}
\end{figure*}

Figure \ref{physics_networks} depicts the collaboration networks of all four classes from Schools 1 and 2. The networks show four types of ties based on whether students display strong or weak relations with others, who at the same time might or not enjoy academic prestige in physics. Beside, the network depiction illustrates grades in shades of green, as well as degree centrality as the nodes' size. 

A first look of these networks allows a visual representation of the information contained on Table \ref{Tabla2}, where classes Sch1-A and B have low network density, represented by the reduced number of ties compared to School 2. Plus, there are differences between groups in School 1, as in class A we observe an abundance of social relations of diverse nature spanning most of the students, whereas class B seems to be governed by strong ties with prestige (blue links) and with many isolated members. In this last group, connected students with high centrality appeared with the highest grades, differently from Sch1-A, where good grades are not an exclusive attribute of central individuals. Finally, both networks on School 2 show an even distribution of good grades, while their ties are mainly strong.   
 
 \begin{figure*}[!htbp]
\centering
  \includegraphics[width=1\textwidth]{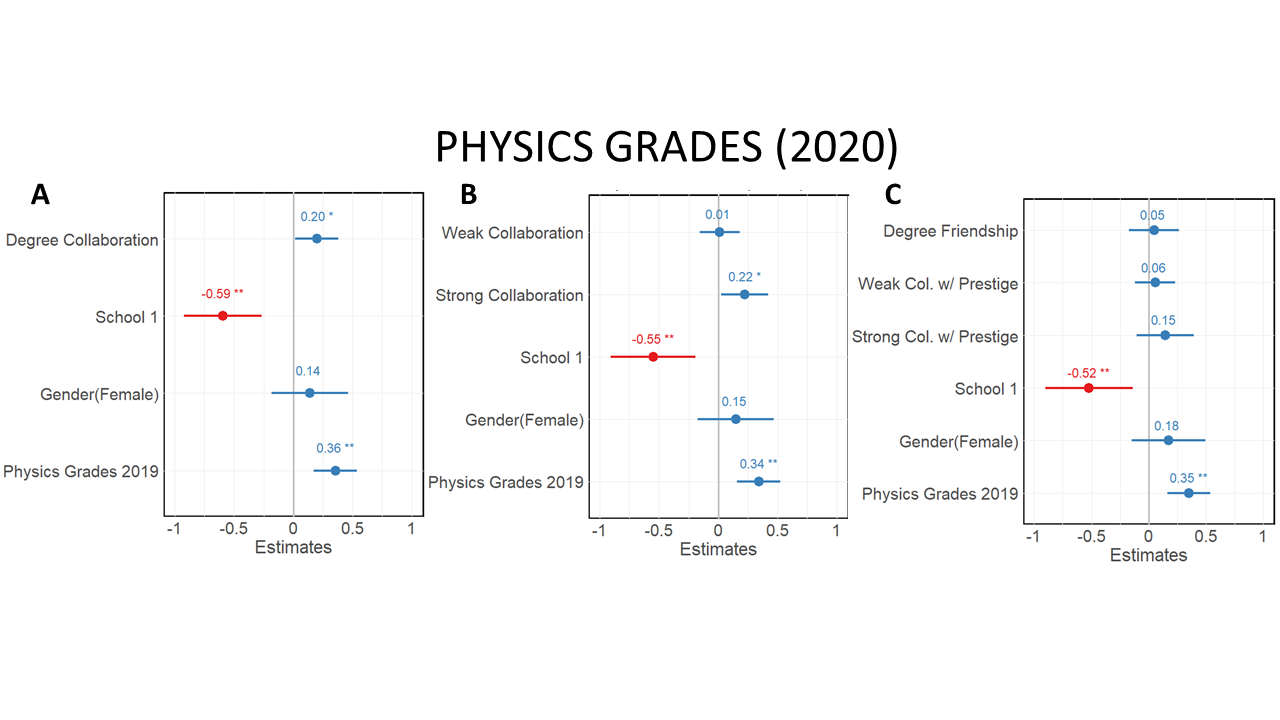}
  \caption{Graphic depiction of OLS multiple linear regression models for physics grades regressed on collaborative variables and confounding variables. Positive coefficients are depicted in blue, while red coefficients indicate negative effects. Note: * $p < .05$ y ** $p < .01$.\label{mod_phys}}
\end{figure*}

The multiple regression models fitted to predict physics grades are shown on Figure \ref{mod_phys}. On model A we used degree centrality of the collaboration network as the main predictor, yielding a positive coefficient (.2, $p < .05$) thus, suggesting that more working partners is associated with success in physics across schools. When differentiating the effects of types of collaboration, model B shows a null effect for weak ties, while having strong relationships indicates a positive predictive value over physics performance (.22, $p < .05$). When adding academic prestige, model C shows the gains of working alongside others who are perceived as good physics students, though this coefficient is not significant. Finally, it is worth highlighting the predictive value of prior grades on physics across all three models, and the mean difference between institutions in favor of School 2.

\section{Methodology: Quasi-Experimental Design}
\subsection{Research Context}
To test the stability of the findings presented in figure \ref{mod_phys} (RQ2), we collected data on four courses from School 2 during the next academic year. To test whether the collaborative gains are observed among student groups based on friendship or randomly assigned (RQ3), we devised a network intervention on two of the four classes from School 2 during two semesters in 2021 (SEM 1 and SEM 2). At the beginning of the academic year (March 2021), students on the experimental conditions ($N_{exp}=46, 50\%$ Female) were randomly assigned to working groups, whilst participants on the control ($N_{cont}=44, 56\%$ Female) condition (2 classes) had the liberty to decide their working teams. This organizational decision led to an initial average of 2 declared friends per group in the control condition, versus an average of 0.5 within-group friends in the experimental courses. The rationale for designing this intervention is grounded on the accessibility of strong and weak collaboration under both experimental conditions and its potential effects over grades. If the results from the exploratory study hold, one would expect the control group to perform better given that students are set to collaborate with their friends (i.e., strong ties), whereas students in the experimental conditions are weakly tied with their working peers. Plus, at the beginning of semester 2, groups in the experimental condition had the liberty to change their working teams, to which we observed 31 (67.4\%) and 13 (30\%) students who decided to change groups in the experimental and control condition respectively. At the end of semester 2, the teacher reported     

Each of the four classes addressed active learning methodologies with group-activities to cover the physics curriculum. Two activities per semester were designed, where students actively engaged with the content, for instance to discuss and analyze conceptual errors in others' responses, generate posters and even design practical experiences to show physics phenomena. Additionally, during the year 2021, and due to a consistent decrease in COVID-19 cases along with a nation-wide immunization policy, schools offered students the possibility to slowly return to face-to-face classes. On this stage, a portion of students in School 2 had the opportunity to participate on face-to-face sessions, while the rest of the class assisted online. We had no information on the average percentage of students per group that attended the school, nor the ones that participated remotely, yet, the teacher reported that by the end of the academic year, the vast majority of the students in each class attended the physics sessions in-person. On this hybrid teaching modality students voluntarily decide whether to attend in-person or online, with the chance to participate in the class in each of these modalities every other week. These contextual conditions are summarized on table \ref{Table4}. 

\begin{table}
\begin{center}
  \caption{Quasi-experimental design and interventions: Longitudinal Analysis \label{Table4}}
    \centering
 \begin{tabular}{p{1.5cm}c p{1.5cm}p{1.5cm}p{2cm}}
\hline\hline
Group  & Subjects & \multicolumn{2}{l}{Grouping Mechanism} & Teaching Modality\\ [-1.8ex] 
& & (SEM$^{*}$ 1) & (SEM$^{*}$ 2) & \\
\hline 
Control & \centering 44   & Affinity & Affinity & Act. Learning \& Hybrid \\
Experimental  & \centering 46   & Random & Affinity &Act. Learning \& Hybrid\\
\hline
\multicolumn{4}{r}{$^*$: Semester} 
\end{tabular}
\end{center}
\end{table}

\subsection{Data Collection and Analysis}
Similar to our experience on the exploratory phase of this study, we collected data on students' collaborative ties, friendship and perceptions of academic proficiency in the course at the end of semester 1 (Activity 1), and two times during semester 2 (Activities 2 and 3). These surveys were administered by the physics teacher via online. in each wave of data collection, and by following the method depicted in figure \ref{methodology}, we constructed the same variables described in \ref{Tabla3}. Similarly, collected data yielded three directed and binary networks. Table \ref{Tabla5} summarized whole network measures (i.e., nodes, ties, average degree and density). Along with this, we gathered information on students' gender, and performance in each of the activities, which came from assessing the structure, organization, language and theory presented on students' group work. 

\begin{table*}[t!]
\begin{center}
  \caption{Descriptive statistics of directed networks on Activities 1, 2 and 3: Friendship, Physics Prestige, \& Collaboration in Physics.\label{Tabla5}}
  \centering
 \begin{tabular}{lcccccccccc}
\hline\hline
\multicolumn{11}{c}{Activity 1}\\
\hline
           &          & \multicolumn{3}{c}{Friendship}   & \multicolumn{3}{c}{Physics Prestige} & \multicolumn{3}{c}{Collaboration on Physics} \\
Group     & Nodes & Ties & Degree$^*$ & Density        & Ties & Degree$^*$ & Density              & Ties & Degree$^*$ & Density\\
\hline 
1-A &	24	&	117	&	4.88	&	0.21	&	147	&	6.13	&	0.27	&	134	&	5.58	&	0.24
 \\
1-B &	24	&	175	&	7.29	&	0.32	&	127	&	5.29	&	0.23	&	113	&	4.71	&	0.21
  \\
2-A &	22	&	125	&	5.68	&	0.27	&	108	&	4.91	&	0.23	&	71	&	3.22	&	0.15
 \\ 
2-B &	20	&	117	&	5.85	&	0.31	&	101	&	5.05	&	0.27	&	57	&	2.85	&	0.15
 \\
\hline
\multicolumn{11}{c}{Activity 2}\\
\hline
          &          & \multicolumn{3}{c}{Friendship}   & \multicolumn{3}{c}{Physics Prestige} & \multicolumn{3}{c}{Collaboration on Physics} \\
Group     & Nodes & Ties & Degree$^*$ & Density        & Ties & Degree$^*$ & Density              & Ties & Degree$^*$ & Density\\
\hline 
1-A &	24	&	220	&	9.17	&	0.4	&	144	&	6	&	0.26	&	131	&	5.46	&	0.24
  \\
1-B &	24	&	193	&	8.04	&	0.35	&	165	&	6.88	&	0.299	&	89	&	3.71	&	0.16
 \\
2-A &	22	&	188	&	8.55	&	0.41	&	134	&	6.09	&	0.29	&	64	&	2.91	&	0.14
  \\ 
2-B &	20	&	188	&	9.4	&	0.5	&	109	&	5.45	&	0.29	&	82	&	4.1	&	0.22
 \\
 \hline
\multicolumn{11}{c}{Activity 3}\\
\hline
          &          & \multicolumn{3}{c}{Friendship}   & \multicolumn{3}{c}{Physics Prestige} & \multicolumn{3}{c}{Collaboration on Physics} \\
Group     & Nodes & Ties & Degree$^*$ & Density        & Ties & Degree$^*$ & Density              & Ties & Degree$^*$ & Density\\
\hline 
1-A &	24	&	198	&	8.25	&	0.34	&	181	&	7.54	&	0.328	&	113	&	4.71	&	0.21
 \\
1-B &	24	&	196	&	8.52	&	0.39	&	160	&	6.67	&	0.29	&	101	&	4.21	&	0.18
 \\
2-A &	22	&	196	&	8.91	&	0.42	&	119	&	5.41	&	0.258	&	72	&	3.27	&	0.16
 \\ 
2-B &	20	&	151	&	7.55	&	0.4	&	112	&	5.6	&	0.3	&	74	&	3.7	&	0.2
 \\
\hline
\multicolumn{11}{l}{$^*$: Average number of incoming plus outgoing ties per student in the class.} 
\end{tabular}
\end{center}
\end{table*}

To find evidence in light of RQs 2 and 3 (Are the academic gains of collaborating with friends and those who enjoy academic prestige stable in time throughout an academic year in high school physics? \& Are these changes similar for student groups based on friendship or randomly assigned? \& Are there differences in the academic gains of different types of collaborative ties for student groups based on friendship or randomly assigned in high school physics?), we fitted OLS multiple linear regression models for each wave of data collection. Here, we regressed physics grades upon collaboration variables (see Table \ref{Tabla3}) as the main predictors. With this, we sought to compare the regression coefficients across activities (models) as evidence of change in the collaborative effect over physics grades. We further checked the robustness of our findings by following \textit{difference-and-difference} \cite{SCHWERDT20203} procedure and tested whether the observed differences in the collaborative effects hold (see Supplementary Material). We found no contradicting evidence from this analysis. In the following section we show the results and related evidence obtained from the analysis.

\section{Results: Quasi-Experimental Design}
\begin{figure*}[!htbp]
\centering
  \includegraphics[width=0.7\textwidth]{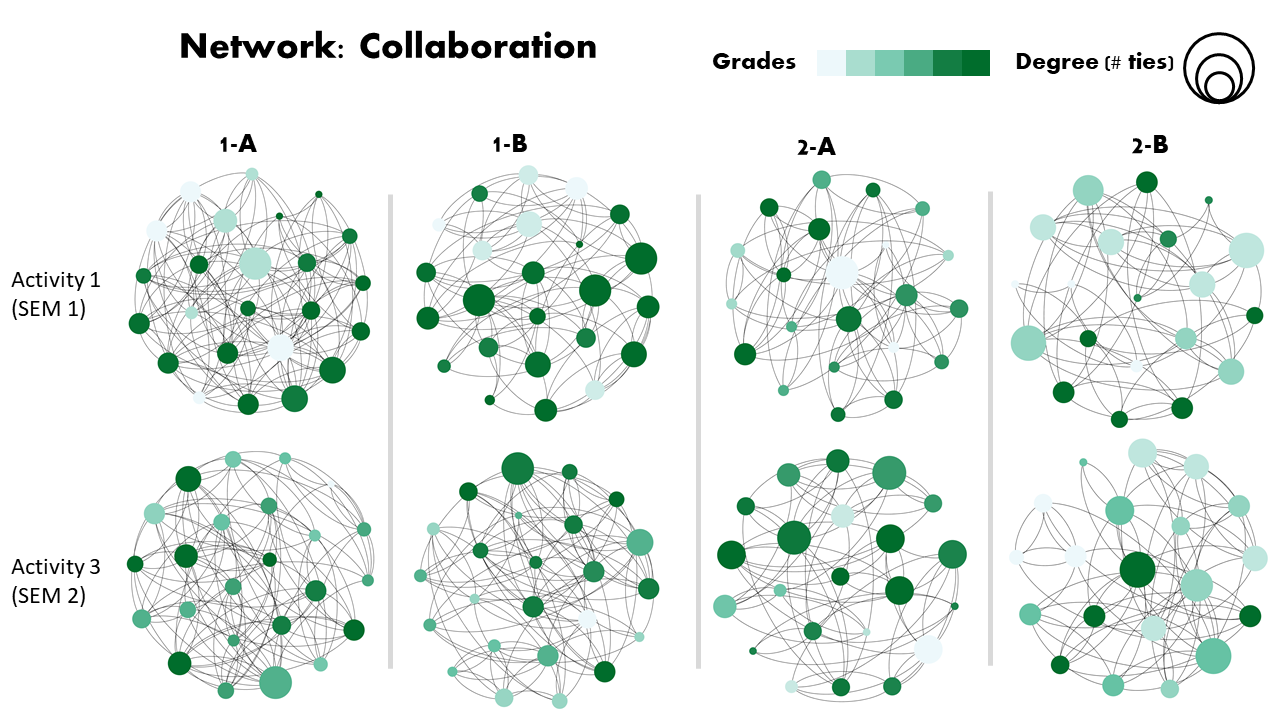}
  \caption{Collaboration network on Activities 1 and 3 for each class. Node colors represents academic performance; size shows degree centrality –number of collaborative ties. \label{n1}}
\end{figure*}

\begin{figure*}[!htbp]
\centering
  \includegraphics[width=0.7\textwidth]{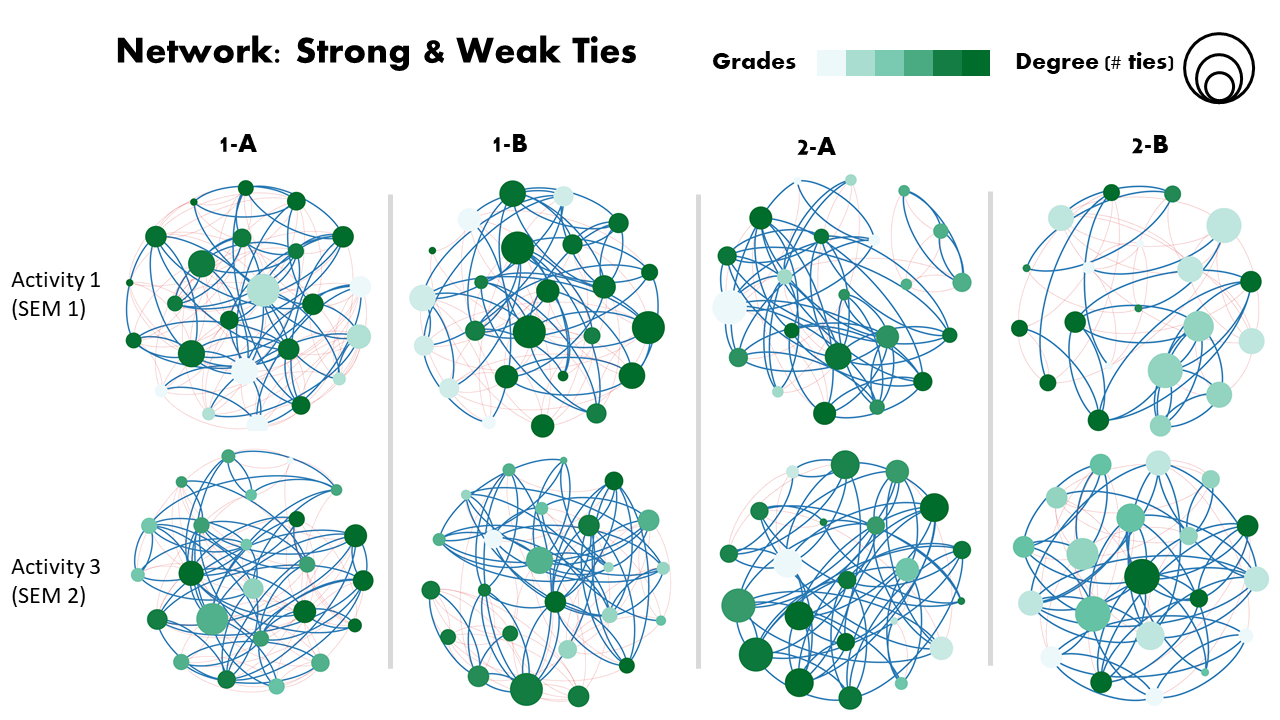}
  \caption{Strong and weak collaboration network on Activities 1 and 3 for each class. Node colors represents academic performance; size shows degree centrality –number of collaborative ties; edges' colors indicate strong ties (blue) and weak ties (red). \label{n2}}
\end{figure*}

\begin{figure*}[!htbp]
\centering
  \includegraphics[width=0.7\textwidth]{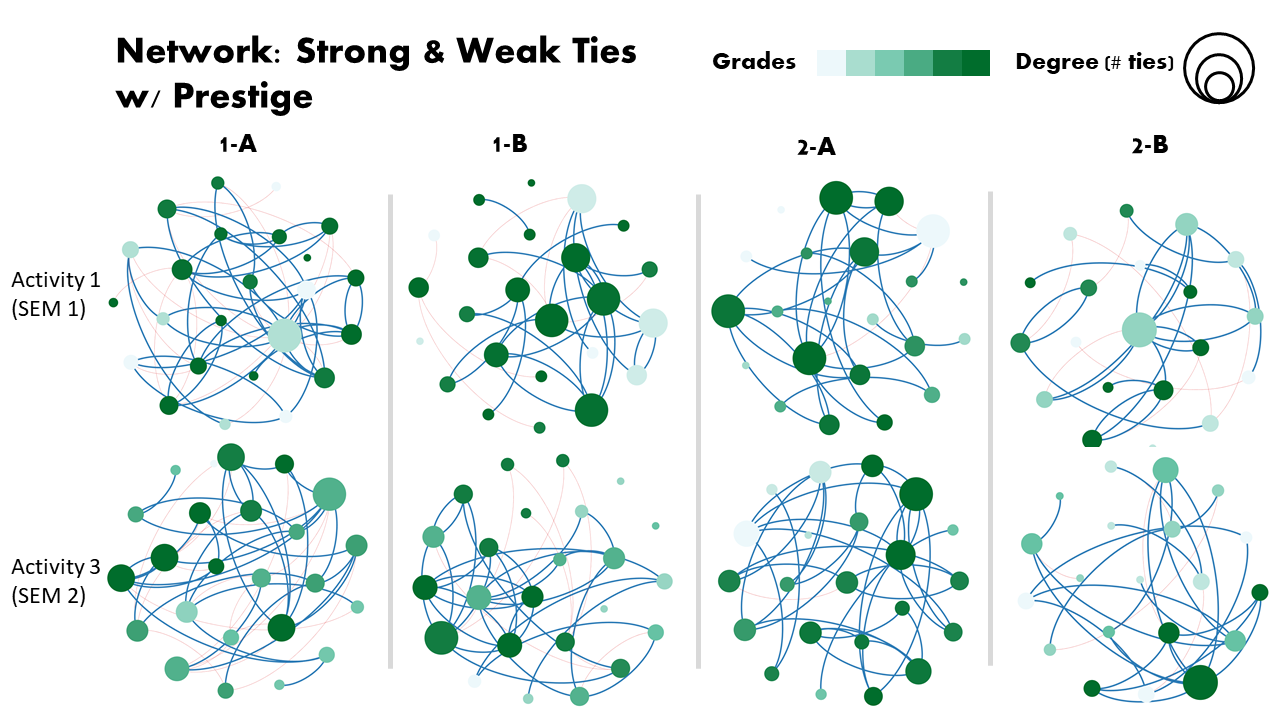}
  \caption{Strong and weak collaboration network with academic prestige on Activities 1 and 3 for each class. Node colors represents academic performance; size shows degree centrality –number of collaborative ties; edges' colors indicate strong ties (blue) and weak ties (red) with proficient  students.\label{n3}}
\end{figure*}

First, figures \ref{n1}, \ref{n2} and \ref{n3} depict students' collaboration networks on activities 1 (end of SEM1) and 3 (end of SEM2), for each of the four classes. Figure \ref{n1} shows the network of collaboration ties (degree), while figures \ref{n2} and \ref{n3} depict respectively, the network of strong and weak collaboration ties, and strong and weak collaboration ties among academically prestigious peers in physics. In all the figures, the size of the nodes represents degree centrality, while darker green indicates higher grades. From a first look into the networks it is noticeable how the network density - number of ties - diminishes as one moves from figure \ref{n1} to \ref{n3}- thus reflecting the restrictions imposed onto strong and weak ties, plus the added dimension of prestige. Further, the location of dark green (i.e., higher grades) and larger nodes (i.e., higher degree centrality ) changes from Activity 1 to Activity 2, thus suggesting a dynamic and complex pattern of relationships in each class.  

\begin{table*}[!htbp]  
  \caption{OLS multiple regression models of physics grades regressed on collaboration predictors and controlling for confounding variables.} 
  \label{table_models2021} 
\scriptsize 
\begin{tabular}{@{\extracolsep{-1pt}}lccccccccc} 
\\[-1.8ex]\hline 
\hline \\[-1.8ex] 
 & \multicolumn{9}{c}{\textit{Dependent variable:}} \\ 
\cline{2-10} 
\\[-1.8ex] & \multicolumn{9}{c}{Physics Grades)} \\ 
\\[-1.8ex] & (1. Act.1) & (2. Act. 2) & (3. Act. 3) & (4. Act. 1) & (5. Act. 2) & (6. Act. 3) & (7. Act. 1) & (8. Act. 2) & (9. Act. 3)\\ 
\hline \\[-1.8ex] 
 Collaboration (Degree) & $-$0.300$^{***}$ & $-$0.123 & 0.067 &  &  &  &  &  &  \\ 
  & (0.102) & (0.121) & (0.105) &  &  &  &  &  &  \\ 
  & & & & & & & & & \\ 
 Weak Collaboration &  &  &  & $-$0.294$^{**}$ & $-$0.213$^{*}$ & $-$0.160 &  &  &  \\ 
  &  &  &  & (0.120) & (0.115) & (0.122) &  &  &  \\ 
  & & & & & & & & & \\ 
 Strong Collaboration &  &  &  & $-$0.256$^{*}$ & $-$0.056 & 0.393$^{**}$ &  &  &  \\ 
  &  &  &  & (0.131) & (0.166) & (0.157) &  &  &  \\ 
  & & & & & & & & & \\ 
 Weak Col. w/ Prestige &  &  &  &  &  &  & $-$0.092 & 0.157 & 0.071 \\ 
  &  &  &  &  &  &  & (0.112) & (0.111) & (0.112) \\ 
  & & & & & & & & & \\ 
 Strong Col. w/ Prestige &  &  &  &  &  &  & $-$0.215 & 0.190 & 0.464$^{***}$ \\ 
  &  &  &  &  &  &  & (0.145) & (0.145) & (0.128) \\ 
  & & & & & & & & & \\ 
 Control Group & $-$0.129 & $-$0.145 & $-$0.046 & $-$0.264 & $-$0.193 & $-$0.168 & $-$0.186 & 0.163 & $-$0.114 \\ 
  & (0.203) & (0.243) & (0.202) & (0.228) & (0.242) & (0.201) & (0.224) & (0.236) & (0.202) \\ 
  & & & & & & & & & \\ 
 Gender (Female) & 0.457$^{*}$ & $-$0.750$^{***}$ & $-$0.613$^{***}$ & 0.472$^{*}$ & $-$0.783$^{***}$ & $-$0.518$^{**}$ & 0.540$^{**}$ & $-$0.658$^{***}$ & $-$0.470$^{**}$ \\ 
  & (0.236) & (0.229) & (0.222) & (0.238) & (0.229) & (0.214) & (0.244) & (0.227) & (0.212) \\ 
  & & & & & & & & & \\ 
 Scores (Prior) & 0.009 & $-$0.009 & 0.205$^{*}$ & 0.033 & $-$0.039 & 0.193$^{*}$ & 0.068 & $-$0.031 & 0.076 \\ 
  & (0.124) & (0.120) & (0.116) & (0.124) & (0.121) & (0.110) & (0.129) & (0.120) & (0.115) \\ 
  & & & & & & & & & \\ 
 Physics Prestige & 0.294$^{**}$ & 0.168 & $-$0.063 & 0.307$^{**}$ & 0.165 & $-$0.057 & 0.396$^{**}$ & 0.057 & $-$0.132 \\ 
  & (0.132) & (0.129) & (0.111) & (0.138) & (0.129) & (0.103) & (0.159) & (0.138) & (0.104) \\ 
  & & & & & & & & & \\
  Friendship (Degree) & 0.044 & 0.029 & 0.051 & 0.032 & $-$0.048 & $-$0.328$^{**}$ & 0.029 & $-$0.010 & $-$0.180 \\ 
  & (0.119) & (0.114) & (0.109) & (0.132) & (0.158) & (0.154) & (0.127) & (0.131) & (0.135) \\ 
  & & & & & & & & & \\ 
 Constant & $-$0.148 & 0.424$^{**}$ & 0.310$^{*}$ & $-$0.085 & 0.464$^{**}$ & 0.327$^{*}$ & $-$0.157 & 0.223 & 0.278 \\ 
  & (0.185) & (0.187) & (0.178) & (0.197) & (0.186) & (0.174) & (0.196) & (0.183) & (0.174) \\ 
  & & & & & & & & & \\ 
\hline \\[-1.8ex] 
Observations & 90 & 90 & 90 & 90 & 90 & 90 & 90 & 90 & 90 \\ 
R$^{2}$ & 0.193 & 0.203 & 0.216 & 0.194 & 0.225 & 0.315 & 0.137 & 0.230 & 0.330 \\ 
Adjusted R$^{2}$ & 0.134 & 0.145 & 0.159 & 0.125 & 0.159 & 0.256 & 0.063 & 0.164 & 0.273 \\ 
Residual Std. Error & 0.930 (df = 83) & 0.925 (df = 83) & 0.917 (df = 83) & 0.936 (df = 82) & 0.917 (df = 82) & 0.862 (df = 82) & 0.968 (df = 82) & 0.914 (df = 82) & 0.852 (df = 82) \\ 
F Statistic & 3.300$^{***}$ & 3.514$^{***}$ & 3.814$^{***}$ & 2.811$^{**}$  & 3.407$^{***}$  & 5.384$^{***}$  & 1.861$^{*}$  & 3.491$^{***}$ & 5.781$^{***}$ \\ 
 & (df = 6; 83) & (df = 6; 83) & (df = 6; 83) & (df = 7; 82) &(df = 7; 82) & (df = 7; 82) & (df = 7; 82) & (df = 7; 82) & (df = 7; 82) \\
\hline 
\hline \\[-1.8ex] 
\textit{Note:}  & \multicolumn{9}{r}{$^{*}$p$<$0.1; $^{**}$p$<$0.05; $^{***}$p$<$0.01} \\ 
\end{tabular} 
\end{table*}

OLS multiple regression models for predicting physics grades in each activity are shown in table \ref{table_models2021}. Models 1, 2 and 3 contain degree centrality in the collaboration network as the main predictor. Models 4, 5 and 6 use the number of strong and weak collaborative ties as the main predictor, whereas for models 7, 8 and 9, strong and weak ties with prestigious others as the key independent variable. 
The first three models show that at the beginning of the academic year the number of collaborative ties yields a negative effect over grades, but by Activity 3 such effect got closer to zero. The effects of strong collaboration (i.e., working ties with friends) however, transition from negative at the end of SEM1 to positive by the end of the year (models 4, 5 and 6). We observe basically no change for working with acquainted peers on models 4, 5 and 6. Finally, the inclusion of physics prestige aligns with the previous results, as working with either friends or acquaintances increases over time. This effect is stronger for those who work with friends that enjoy academic prestige. 

\begin{figure*}[!htbp]
\centering
  \includegraphics[width=1.05\textwidth]{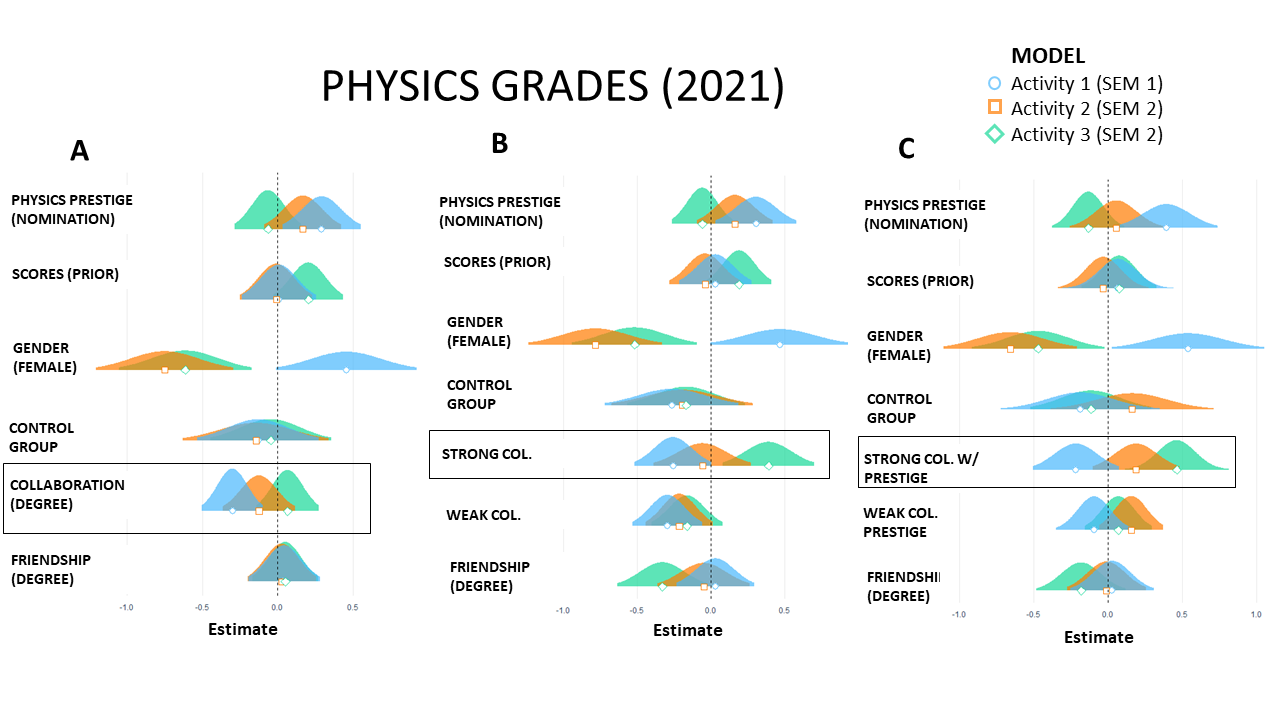}
  \caption{Graphic depiction of OLS multiple linear regression coefficient comparison between Activity 1 at the end of Semester 1 (light blue); Activity 2 at the beginning of Semester 2 (orange); and Activity 3 at the end of Semester 2 (light blue). \label{comparative models 2021}}
\end{figure*}

Figure \ref{comparative models 2021} depicts the comparison between the regression coefficients shown in table \ref{table_models2021}. Here, it is possible to observe the different effects of collaboration variables in performance across Activities 1 (light blue), Activity 2 (orange) and Activity 3 (light green). The squared coefficients on figure \ref{comparative models 2021} evidence the evolving effect of classroom collaboration towards effectiveness throughout two academic semesters. Note that from Activity 1 towards Activity 2 a percentage of students under both experimental and control conditions decided to pair with their friends, thus increasing their chances to higher scores through consistently working with friends (i.e., strong ties). Interestingly, physics prestige loses power in predicting grades across activities, thus suggesting that achieving high performance is no longer an exclusive reward for those who are deemed as proficient in high school physics. 

From table \ref{table_models2021} the experimental group (i.e., random grouping mechanism) in general got higher scores than the control group (i.e., friendship grouping mechanism), yet these differences are not significant, and indicate that both types of group configurations (random and friendship-based) afford similar levels of academic achievement. Moreover, on average, females got better grades than males on Activity 1, but performed worse during the second semester (Activities 2 and 3).

\section{Discussion}

\subsection{Exploratory Study}
The multiple regression models fitted for the exploratory study show the expected results of having multiple working peers in the classroom, which align with the benefits of social integration on education \cite{Tinto}. The observed learning gains are also consistent with sociocultural theory \cite{vyg} and the principles of social capital \cite{zhao2010}, in the sense that social relationships enable access to information and ease collective learning, in this context, pertaining to physics. Nonetheless, when examining the academic effects of different collaborative relationships, strong friendship ties showed an advantage compared to weak ties observed among those who do not declare friendship ties, which is evidence to answer RQ1. One could attribute such effect to the type of information needed to successfully address activities in each school, following the evidence from Hansen \cite{hansen} and Granovetter \cite{Granovetter}, with complex ideas being better disseminated and learned through strong ties, because individuals on strongly connected networks are proxies of collective trust and commitment, with shared common forms of communication. 

Alternatively, if the information-based approach fails to explain the effect of strong ties on physics grades, we could appeal to the contrasting nature of the learning activities administered in School 1 and 2. Here, the effectiveness of strong relationships implies the adoption of strategic behaviors from working with friends to overcome the academic challenges embedded in each task. Plus, it is possible that remote teaching accentuates existing students' social networks due to the costs of expanding one's ties via ICT's mediated communication. In simple terms, collaboration with friends implies a lesser effort than embarking on new working ties through remote activities, compared with face-to-face interactions, where verbal and non-verbal cues, along with physical proximity reduce the costs and uncertainties of working with less known peers.

Results show clear differences in the density of collaboration and friendship ties between both schools. This contrasting scenario might be attributable to the socioeconomic and cultural characteristics found on both student populations. Even though we do not count with information about students' social or economic capital, schools' geographical locations in rural and urban areas allow us to imply certain comparative advantages relative to the access to internet and communication technologies, which might have facilitated, or otherwise hindered students' interactions for miscellaneous purposes. According to the literature on online education and ICTs, success depends highly on students' accessibility and their readiness to use digital tools \cite{Hung2010, Kebritchi2017}. Consequently, one could think that students from School 1 (rural) experienced higher limitations on accessing internet and interacting with ICTs, yielding to less ICTs' mediated communication and interactions compared to School 2, and thus resulting in lower network density in both friendship and prestige networks. Higher exposure to others' comments, behaviors and grades due to easier ICT accessibility becomes a form of social information that is reflected in the higher average of ties observed in the friendship and prestige networks on School 2.     

Besides the difference in the number of working ties on Schools 1 and 2, we found no significant interactions between the schools and the collaborative variables to support the claim that context and teaching methodologies might have led to diverse academic gains. The lack of differences across schools might be attributed to distinctive social processes enacted in the face of the teaching strategies observed in each learning context. As both schools put in place a traditional (School 1) and active (School 2) physics classrooms, it is reasonable to think that the ways in which students worked together to meet their academic goals differed between both teaching methodologies. Even though students on School 1 might have addressed the task in online groups, traditional well-structured problems are not characterized by requiring high levels of interdependence and effective collective coordination \cite{Johnson_Johnson, Steiner1966, pulgar_2021}, and consequently, partnerships with one or two closest friends would be enough to access and develop appropriate knowledge to solve the activities. Conversely, the open-ended nature of the activities observed in School 2 should have encouraged student groups higher levels of coordination and communication with their two or three group members. We resort to this interpretation given the lack of insight knowledge on the collective processes these activities encouraged. Because in all four courses the vast majority of the collaborative ties are observed between friends, it is clear that students were able to identify effective working partners to meet the academic goals in each learning scenario.

\subsection{Quasi-Experimental Design}

As shown, the effects of collaboration over physics grades evolve along with external social re-configurations. Throughout the academic year, collaboration becomes more effective in fostering good performance in high school physics, particularly for those who form partnerships with friends and those perceived as good physics students, and the response to RQ2. The dynamic effects of the different collaborative ties captured throughout the three activities could be a consequence of both structural conditions and personal processes. Schools and classroom organization, as in the hybrid modality with half the students attending the classroom while the other half remained online during the first semester, presumably slowed down groups' coordination and communication, and likely conducive towards negative effect over physics grades. In the control groups for instance, it is likely that students found themselves at the beginning of the year teaming up with friends who were attending the sessions remotely, with the alternative configuration (i.e., most friends connected online) a likely scenario as well. In either case, developing group-level strategies between face-to-face and remote communication is non-trivial due to the difficulties for online members to catch up with face-to-face conversation, and vice-versa. Differently, students in the experimental group faced an additional challenge to coordinate their groups' work on semester 1, either in-person and/or remotely with less known classmates. Even though this added complexity, the experimental and control groups performed at the same level during first and second semesters. 

Based on our results, there are no performance differences between both experimental conditions (RQ3). The lack of performance differences between student groups from control and experimental conditions during the first semester might also be caused by the negative symptoms of working with close friends within a highly cohesive network. From an effort perspective, a working environment constructed among friends could hinder responsibility and individual accountability \cite{le_collaborative_learning_practices}, or alternatively, high cohesion could lead to overestimating the individual and collective capabilities to perform a particular task \cite{wise,park}. Conversely, and accounting for all the challenges associated with coordination and building working ties, groups in the experimental condition could have overcome such initial limitations by assembling effort in response to this new working scenario. Besides, it is possible that students hold a strong belief regarding their role as learners in a traditional physics classroom, where the attention goes to individual performance via well-structured activities, rather than collective performance. Such performance-driven identity might also be caused by the learning environment experienced in other disciplines. Therefore, facing open-ended and group activities could have required a level of adaptation that teachers must include in the daily practices.  

Further into the year, and after having the chance to reconfigure their groups, collaboration became increasingly more effective in affording better grades, particularly for those who declare working ties with friends and good students. A larger number of new groups were formed in the experimental courses, presumably due to participants seeking out to form teams with close friends and proficient classmates. On top of that, at the end of the second semester, a vast majority of students were present in the classroom. Consequently, the combination of face-to-face collaboration with friends in a learning environment already known for requiring non-traditional group tasks, could have eased the development of adequate group-level processes, thus boosting teams' effectiveness in the classroom. Besides the managerial decision and liberties experienced during the second semester, is it also likely that students in general optimized their working relationships by the end of the year. This implies either that participants strengthen their effective social ties, or rather discarded unfruitful ones to work alongside more academically-oriented peers, which constitutes an opportunity to develop critical collaborative abilities. This is observed in the effects of strong and weak ties with prestigious peers in physics, because even in the case of weak relationships, there was a value added at the end of the year for interacting with good students outside the personal network of friends.  

Interestingly, students who are perceived as proficient in physics classrooms scored significantly better only on Activity 1. During Activity 1 it is possible that group members reproduced the group-level strategies enacted on traditional physics problems, resulting in more frequent attention to prestigious members within each group and across working teams \cite{Pulgar19}. On activities held during the second semester however, the effect of such academic hierarchies is reduced to zero and even becomes negative. The contrasting academic effects for physics prestige and collaboration with good students is evidence that good results are associated with collaboration with these individuals rather than their isolated knowledge and skills. This supports our prior interpretations that groups built up better collective processes while getting accustom to the nature and requirements of additive tasks on semester 2, and with performance opportunities that extend to all members who engaged in collaboration, and beyond the socially constructed hierarchies defined by perceptions of physics proficiency. 

Finally, the collaboration networks measured throughout the year for the quasi-experiment are notably more dense than their similar networks mapped during the remote teaching modality. As mentioned, this is not surprising as face-to-face interactions seem to afford higher chances for connecting classmates compared to digitally mediated communications, under the assumption that students indeed utilize formal means to relate with each other (e.g., chats and forums). Additionally, the motivation to attend face-to-face sessions at the school increased students' social engagement considerably from the first to the second semester, and by the end of the year teachers witnessed a vast majority of students were in the classroom participating in the group activities. 


\subsection{Teaching Implications, Future Questions and Limitations}
The similar performance observed on both control and experimental groups, specially on Activity 1, is an encouraging sing for using random groups in physics classrooms. Even though the collaborative effects during the first semester were not what we would have expected, it is possible that consistency brings the desired academic gains. Accordingly, future research should explore in detail whether continues work in randomly assigned groups in physics education fosters appropriate coordination for learning and good performance. Plus, interesting processes such as inter-group collaboration, or more complex mechanisms like brokerage are future avenues for research, particularly given the creative orientation in today's education \cite{sawyer_edu_innovation, Pulgar_TSC}. Accordingly, managerial decisions such as forming random or friend-based groups could have encouraged students to reach other groups based on the likelihood of having their friends outside their groups. In randomly assigned groups, having friendship ties with other team members could foster first, more frequent between-group interactions and the possibility for students to bridge structural holes \cite{Pulgar_TSC}, a relevant process for creativity in physics classrooms \cite{Pulgar19} and other fields \cite{Burt2004}. Plus, experiencing the need to develop working ties with less known team members becomes an opportunity to extend and fortify one's social network, because effective working ties might evolve into friendship, whereas it becomes also possible to discard ineffective collaborative relationships \cite{network_intervention, Smirnov2017}. 

Friend-based groups seemed to ease collective coordination and the consequent advantage of social relationships. This is a positive sign in the pursuit of social competencies in the classroom, which has been found limited in some student populations \cite{le_collaborative_learning_practices}. Additionally, it is important to include collaborative guidelines for effective performance, especially in students who are not accustomed to learning processes associated with positive interdependence, individual accountability, coordination and decision-making \cite{Johnson_Johnson, pulgar_2021}. This becomes critical in traditional learning contexts where the nature of the scientific knowledge is grounded in well-structured and close-ended problems, and thus fortifies a certain way of performing in physics classrooms that differs from real-world scientific and professional endeavours.  

Finally, both documented research experiences lack information on: internet connectivity; accessibility and teacher and student training on ICTs, given the major role these play in remote education; and in-depth characterization of interactions between the participants physically present in the classroom and those attending remotely in year 2021. Besides, we recognize that the conducted study does not account for students' personal experience during the transition from face-to-face to online instruction, nor the difficulties experienced by teachers on such transitions. For instance, data on teachers' behaviors and strategies to guide the social processes on both remote and hybrid teaching modality could clarify underlying difficulties or effective group-level mechanisms in the classroom. Here, qualitative information and analytical tools might have provided valuable information to either support or reject the interpretations described in the paragraphs above. Furthermore, the study pretends to encourage the research and teaching community to reflect and  pursue new studies and interventions on the interplay between students' networks, teaching methods and learning. This avenue of research is promising, particularly since social and collaborative skills are nowadays necessary for human development and to face the complex challenges of the XXI Century \cite{Bao2019, sawyer_edu_innovation}. 

\section{Conclusions}
We found evidence to answer the three research questions that motivate this study. First, friendship-based working ties, or what we have defined as strong ties are more effective in fostering grades than working with less known peers in the classroom during fully remote teaching modalities. These results hold on both traditional and group-based active learning methodologies. Even though the learning conditions changed from one year to the next giving students flexibility to attend schools, we found that the effectiveness of strong ties gain strength with continuity and persistence, with end of the year activities benefiting more from friendship-based working ties. These gains, presumably come along with better group-level processes and familiarity with unstructured activities. Finally, we found no statistical differences between randomly assigned and friendship-based groups in terms of performance, which is an encouraging sign, meaning that such pedagogical decisions are not necessarily detrimental for students' grades.   

\section*{Acknowledgement(s)}

This material is supported by project FONDECYT 11220385, ANID, Chile; and project 2020217 IF/F, Vicerrecto\'{i}a de Investigaci\'{o}n y Postgrado, Universidad del B\'{i}o B\'{i}o, Concepci\'{o}n, Chile.


\begin{thebibliography}{10}

\bibitem{echeita2012}
G.~Echeita.
\newblock El aprendizaje cooperativo al servicio de una educación de calidad:
  Cooperar para aprender y aprender a cooperar.
\newblock In J.~C. Torrego and A.~Negro, editors, {\em Aprendizaje Cooperativo
  en las aulas: Fundamentos y Recursos para su Implantación}. Alianza
  Editorial, Madrid, España, 2012.

\bibitem{Barkley_ColTech}
E.~Barkley, C.~H. Major, and K.~P. Cross.
\newblock {\em Collaborative learning techniques: A handbook for college
  faculty}.
\newblock Jossey-Bass a Wiley Brand, San Francisco, CA, USA, 2014.

\bibitem{cerda2019}
G.~Cerda, C.~P\'{e}rez, P.~Elipe, J.A. Casas, and R.~Del~Rey.
\newblock Convivencia escolar y su relación con el rendimiento académico en
  alumnado de educación primaria.
\newblock {\em Revista de Psicodidáctica}, 24(1):46--52, 2019.

\bibitem{Bao2019}
L.~Bao and K.~Koenig.
\newblock Physics education research for 21st century learning.
\newblock {\em Disciplinary and Interdisciplinary Science Education Research},
  1(2):1--12, 2019.

\bibitem{pllegrino}
J.W. Pellegrino and M.L. Hilton.
\newblock {\em Education for Life and Work: developing transferable knowledge
  and skills for 21st century}.
\newblock The National Academies Press, New York, 2003.

\bibitem{cabrera_cl_2002}
Alberto Cabrera, Jennifer Crissman, Elena Bernal, Amaury Nora, Patrick
  Terenzini, and Ernest Pascarella.
\newblock Collaborative learning: Its impact on college students' development
  and diversity.
\newblock {\em Journal of College Student Development}, 43(1):20--34, 2002.

\bibitem{sawyer_edu_innovation}
R.K. Sawyer.
\newblock Educating for innovation.
\newblock {\em Thinking Skils and Creativity}, 1(1):41--48, 2006.

\bibitem{Vonderwell2005}
S.~Vonderwell and S.~Zachariah.
\newblock Factors that influence participation in online learning.
\newblock {\em Journal of Research on Technology in Education}, 38:213--230,
  2005.

\bibitem{Traxler2018}
A.~Traxler, A.~Gavrin, and R.~Lindell.
\newblock Networks identify productive forum discussions.
\newblock {\em Physical Review Physics Education Research}, 14:020107, 2018.

\bibitem{panigrahi}
R.~Panigrahi, P.~R. Srivastava, and D.~Sharma.
\newblock Online learning: Adoption, continuance, and learning outcome—a
  review of literature.
\newblock {\em International Journal of Information Management}, 43:1--14,
  2018.

\bibitem{Hung2010}
M.~Hung, C.~Chou, C.~Chen, and Z.~Own.
\newblock Learner readiness for online learning: Scale development and student
  perceptions.
\newblock {\em Computers \& Education}, 55:1080--1090, 2010.

\bibitem{Kebritchi2017}
M.~Kebritchi, A.~Lipschuetz, and L.~Santiague.
\newblock Issues and challenges for teaching successful online courses in
  higher education.
\newblock {\em Journal of Educational Technology Systems}, 46(1):4--29, 2017.

\bibitem{Romiszowski2004}
A.~Romiszowski and R.~Mason.
\newblock Computer-mediated communication.
\newblock In H.~Jonassen, editor, {\em Handbook of research for educational
  communications and technology}. Lawrence Erlbaum, New Jersey, NJ, 2004.

\bibitem{Wise2013}
A.~F. Wise, J.~Speer, F.~Marbouti, and Y.~Hsiao.
\newblock Broadening the notion of participation in online discussions:
  Examining patterns in learners’ online listening behaviors.
\newblock {\em Instructional Science}, 41:323--343, 2013.

\bibitem{le_collaborative_learning_practices}
Ha~Le, Jeroen Janssen, and Theo Wubbels.
\newblock Collaborative learning practices: teacher and student perceived
  obstacles to effective student collaboration.
\newblock {\em Cambridge Journal of Education}, 48(1):113--122, 2018.

\bibitem{pulgar_2021}
J.~Pulgar, V.~Fahler, and A.~Spina.
\newblock Investigating how students collaborate to compose physics problems
  through structured tasks.
\newblock {\em Physical Review Physics Education Research}, 17:010120, 2021.

\bibitem{Pulgar_TSC}
J.~Pulgar.
\newblock Classroom creativity and students' social networks: theoretical and
  practical implications.
\newblock {\em Thinking Skills and Creativity}, In Press.

\bibitem{Grunspan}
D.~Z. Grunspan, B.~L. Wiggins, and S.~M. Goodreau.
\newblock Understanding classrooms through social network analysis: A primer
  for social network analysis in educational research.
\newblock {\em CBE-Life Sciences Education}, 13:167--178, 2014.

\bibitem{Putnik}
G.~Putnik, E.~Costa, C.~Alves, H.~Castro, L.~Varela, and V.~Shah.
\newblock Analyzing the correlation between social network analysis measures
  and performance of students in social network-based engineering education.
\newblock {\em International Journal of Technology and Design Education},
  26:413--437, 2016.

\bibitem{Bruun2013}
J.~Bruun and E.~Brewer.
\newblock Talking and learning physics: predicting future grades from network
  measures and force concept inventory pretests scores.
\newblock {\em Physical Review Physics Education Research}, 9:021109, 2013.

\bibitem{Pulgar19}
J.~Pulgar, C.~Candia, and P.~Leonardi.
\newblock Social networks and academic performance in physics: Undergraduate
  cooperation enhances ill-structured problem elaboration and inhibits
  well-structured problem solving.
\newblock {\em Physical Review Physics Education Research}, 16:010137, 2020.

\bibitem{Brewe_Sawtelle2012}
E.~Brewe, L.~Kramer, and V.~Sawtelle.
\newblock Investigating student communities with network analysis on
  interactions in a physics learning center.
\newblock {\em Physical Review Physics Education Research}, 8:010101, 2012.

\bibitem{Morris2005}
K.~V. Morris, C.~Finnegan, and W.~Sz-Shyan.
\newblock Tracking student behavior, persistence, and achievement in online
  courses.
\newblock {\em Internet and Higher Education}, 8:221--231, 2005.

\bibitem{Dawson2008}
S.~Dawson.
\newblock A study of the relationship between student social networks and sense
  of community.
\newblock {\em Educational Technology \& Society}, 11(3):224--238, 2008.

\bibitem{vyg}
L.~S. Vygotsky.
\newblock {\em Mind in Society: the Development of Higher Psychological
  Processes}.
\newblock Harvard University Press, Cambridge, MA, 1978.

\bibitem{Williams}
E.~A. Williams, J.~P. Zwolak, R.~Dou, and E.~Brewe.
\newblock Linking engagement and performance: The social network analysis
  perspective.
\newblock {\em Physical Review Physics Education Research}, 15, 2019.

\bibitem{Zwolak_Zwolak}
J.P. Zwolak, M.~Zwolak, and E.~Brewe.
\newblock Educational commitment and social networking: The power of informal
  networks.
\newblock {\em Physical Review Physics Education Research}, 14:010131, 2018.

\bibitem{Zwolak}
J.~P. Zwolak, R.~Dou, E.~A. Williams, and E.~Brewe.
\newblock Students’ network integration as a predictor of persistence in
  introductory physics courses.
\newblock {\em Physical Review Physics Education Research}, 13:010113, 2017.

\bibitem{Dou2016}
R.~Dou, E.~Brewe, J.~P. Zwolak, G.~Potvin, E.~A. Williams, and L.~H. Kramer.
\newblock Beyond performance metrics: examining a decrease in students’
  physics self-efficacy through a social network lens.
\newblock {\em Physical Review Physics Education Research}, 12:020124, 2016.

\bibitem{Carrasco2018}
C.~Carrasco, R.~Alarcón, and M.~V. Trianes.
\newblock Adatación y trabajo cooperativo en el alumnado de educación
  primaria desde la percepción del profesorado y la familia.
\newblock {\em Revista de Psicodidáctica}, 23(1):56--62, 2018.

\bibitem{leon2017}
B.~León, S.~Mendo-Lázaro, E.~Felipe-Castaño, M.I. Polo, and
  F.~Fajardo-Bullón.
\newblock Potencia de equipo y aprendizaje cooperativo en el ámbito
  universitario.
\newblock {\em Revista de Psicodidáctica}, 22(1):9--15, 2021.

\bibitem{Burt2004}
R.~S. Burt.
\newblock Structural holes and good ideas.
\newblock {\em American Journal of Sociology}, 110(2):349--399, 2004.

\bibitem{Candia}
C.~Candia, V.~Landaeta-Torres, C.~A. Hidalgo, and C.~Rodriguez-Sickert.
\newblock Strategic reciprocity improves academic performance in public
  elementary school children.
\newblock {\em Preprint arXiv:1909.11713}, 2019.

\bibitem{Granovetter}
M.S. Granovetter.
\newblock The strength of weak ties.
\newblock {\em American Journal of Sociology}, 78(6):1360--1380, 1973.

\bibitem{hansen}
M.~T. Hansen.
\newblock The search-transfer problem: The role of weak ties in sharing
  knowledge across organization subunits.
\newblock {\em Administrative Science Quarterly}, 44(1):82--111, 1999.

\bibitem{dokuka}
S.~Dokuka, D.~Valeeva, and M.~Yudkevich.
\newblock How academic achievement spreads: The role of distinct social
  networks in academic performance diffusion.
\newblock {\em PLOS ONE}, 15(15), 2020.

\bibitem{wise}
S.~Wise.
\newblock Can a team have too much cohesion? \uppercase{T}he dark side to
  network density.
\newblock {\em European Management Journal}, 32:703--711, 2014.

\bibitem{park}
W.~Park, M.~S. Kim, and S.~M. Gully.
\newblock Effect of cohesion on the curvilinear relationship between team
  efficacy and performance.
\newblock {\em Small Group Research}, 48(4):455--481, 2017.

\bibitem{Hrastinski2009}
S.~Hrastinski.
\newblock A theory of online learning as online participation.
\newblock {\em Computers \& Education}, 52:78--82, 2009.

\bibitem{deweber}
B.~De~Wever, T.~Schellens, M.~ValckeH, and H.~Van~Keer.
\newblock Content analysis schemes to analyze transcripts of online
  asynchronous discussion groups: A review.
\newblock {\em Computers \& Education}, 46:6--28, 2006.

\bibitem{dawson2010}
S.~Dawson.
\newblock Seeing the learning community: An exploration of the development of a
  resource for monitoring online student networking.
\newblock {\em British Journal of Educational Technology}, 41, 2010.

\bibitem{traxler2020}
A.~T. Traxler, T.~Suda, E.~Brewe, and K.~Commeford.
\newblock Network positions in active learning environments in physics.
\newblock {\em Physical Review Physics Education Research}, 16:020129, 2020.

\bibitem{Commeford2020}
K.~Commeford, E.~Brewe, and A.~T. Traxler.
\newblock Characterizing active learning environments in physics using network
  analysis and copus observations.
\newblock {\em Preprint arXiv:2008.05325}, 2020.

\bibitem{Pulgar_Sochifi}
J.~Pulgar, C.~R\'{i}os, and Cristian Candia.
\newblock Physics problems and instructional strategies for developing social
  networks in university classrooms.
\newblock {\em Preprint arXiv:1904.02840}, 2019.

\bibitem{Brewe_Kramer2012}
E.~Brewe, L.~H. Kramer, and G.~E. O’Brien.
\newblock Changing participation through the formation of student learning
  communities.
\newblock In {\em AIP Conference Proceedings}, volume 1289, pages 85--88, 2012.

\bibitem{chametzky}
B.~Chametzky.
\newblock Andragogy and engagement in online learning: Tenets and solutions.
\newblock {\em Creative Education}, 5:813--821, 2014.

\bibitem{Luyt2013}
I.~Luyt.
\newblock Bridging spaces: Cross-cultural perspectives on promoting positive
  online learning experiences.
\newblock {\em Journal of Educational Technology Systems}, 42:3--20, 2013.

\bibitem{Niess2013}
M.~Niess and H.~Gillow-Wiles.
\newblock Developing asynchronous online courses: Key instructional strategies
  in a social metacognitive constructivist learning trajectory.
\newblock {\em Journal of Distance Education}, 27, 2013.

\bibitem{gupta2013}
S.~Gupta and R.~Bostrom.
\newblock Research note-an investigation of the appropriation of
  technology-mediated training methods incorporating enactive and collaborative
  learning.
\newblock {\em Information Systems Research}, 24(2):454--469, 2013.

\bibitem{merono2021}
L.~Meroño, A.~Calderón, and J.~Arias-Estero.
\newblock Pedagogía digital y aprendizaje cooperativo: efecto sobre los
  conocimientos tecnológicos y pedagógicos del contenido y el rendimiento
  académico en formación inicial docente.
\newblock {\em Revista de Psicodidáctica}, 26(1):53--61, 2021.

\bibitem{Johnson_Johnson}
D.~W. Johnson, R.~T. Johnson, and E.~J. Holubec.
\newblock {\em Circles of Learning: Cooperation in the Classroom}.
\newblock Interaction, Edina: MN, 1986.

\bibitem{Fortus}
D.~Fortus.
\newblock The importance of learning to make assumptions.
\newblock {\em Science Education}, 93(1):86--108, 2008.

\bibitem{elementary_network_2017}
C.-C. Liu, Y.-C. Chen, and S.-J. Daian~Tai.
\newblock A social network analysis on elementary student engagement in the
  networked creation community.
\newblock {\em Computers \& Education}, 115:114--125, 2017.

\bibitem{lab_collaboration2022}
Meagan Sundstrom, David~G. Wu, Cole Walsh, Ashley~B. Heim, and N.~G. Holmes.
\newblock Examining the effects of lab instruction and gender composition on
  intergroup interaction networks in introductory physics labs.
\newblock {\em Phys. Rev. Phys. Educ. Res.}, 18:010102, Jan 2022.

\bibitem{Steiner1966}
I.~D. Steiner.
\newblock Models for inferring relationships between group size and potential
  productivity.
\newblock {\em Behavioral Science}, 11:273--283, 1966.

\bibitem{teo}
R.~Teodorescu, C.~Bennhold, G.~Feldman, and L.~Medsker.
\newblock New approach to analyzing physics problems: Ataxonomy of introductory
  physics problems.
\newblock {\em Physical Review Physics Education Research}, 9:010103, 2013.

\bibitem{Anderson2001}
L.~W. Anderson and D.R. Krathwohl.
\newblock {\em A taxonomy for learning, teaching, and assessing: A revision of
  Bloom's Taxonomy of Educational Objectives}.
\newblock Addison-Wesley/Longman, New York, USA, 2001.

\bibitem{vignery2020achievement}
Kristel Vignery and Wim Laurier.
\newblock Achievement in student peer networks: A study of the selection
  process, peer effects and student centrality.
\newblock {\em International Journal of Educational Research}, 99:101499, 2020.

\bibitem{stadtfeld2019integration}
Christoph Stadtfeld, Andr{\'a}s V{\"o}r{\"o}s, Timon Elmer, Zs{\'o}fia Boda,
  and Isabel~J Raabe.
\newblock Integration in emerging social networks explains academic failure and
  success.
\newblock {\em Proceedings of the National Academy of Sciences},
  116(3):792--797, 2019.

\bibitem{rizzuto2009s}
Tracey~E Rizzuto, Jared LeDoux, and John~Paul Hatala.
\newblock It’s not just what you know, it’s who you know: Testing a model
  of the relative importance of social networks to academic performance.
\newblock {\em Social Psychology of Education}, 12(2):175--189, 2009.

\bibitem{status_vs_popularity}
Martin~H. Jones and Toby~J. Cooke.
\newblock Social status and wanting popularity: different relationships with
  academic motivation and achievement.
\newblock {\em Social Psychology of Education}, 24:1281--1303, 2021.

\bibitem{Borgatti2013}
S.~P. Borgatti, M.~G. Everett, and J.C. Johnson.
\newblock {\em Analyzing Social Networks}.
\newblock SAGE, Washington, DC, 2013.

\bibitem{Brewe_Bruun16}
E.~Brewe, J.~Bruun, and I.~G. Bearden.
\newblock Using model analysis for multiple choice responses: A new method
  applied to force concept inventory data.
\newblock {\em Physical Review Physics Education Research}, 12:020131, 2016.

\bibitem{carolan}
B.~V. Carolan.
\newblock {\em Social Network Analysis and Education: Theory, Methods and
  Applications}.
\newblock SAGE Publications Inc, 2014.

\bibitem{SCHWERDT20203}
Guido Schwerdt and Ludger Woessmann.
\newblock Chapter 1 - empirical methods in the economics of education.
\newblock In Steve Bradley and Colin Green, editors, {\em The Economics of
  Education (Second Edition)}, pages 3--20. Academic Press, second edition
  edition, 2020.

\bibitem{Tinto}
V.~Tinto.
\newblock Classrooms as communities: Exploring the educational character of
  student persistence.
\newblock {\em Journal of Higher Education}, 68(6):599--623, 1997.

\bibitem{zhao2010}
L.~Zhao, Y.~Lu, B.~Wang, P.~Y. Chau, and L.~Zhang.
\newblock Cultivating the sense of belonging and motivating user participation
  in virtual communities: A social capital perspective.
\newblock {\em International Journal of Information Management},
  32(6):574--588, 2010.

\bibitem{network_intervention}
Zs\'{o}fia Boda, Timon Elmer, Andr\'{a}s V\"{o}r\"{o}s, and Christoph
  Stadtfeld.
\newblock Short-term and long-term effects of a social network intervention on
  friendships among university students.
\newblock {\em Scientific Reports}, 10:2889, 2020.

\bibitem{Smirnov2017}
Ivan Smirnov and Stefan Thurner.
\newblock {Formation of homophily in academic performance: Students change
  their friends rather than performance}.
\newblock {\em PLOS ONE}, 12(8), aug 2017.

\end{thebibliography}

\pagebreak
\clearpage
\section{Supplementary Material}

\begin{table*}[!htbp] \centering 
  \caption{OLS multiple linear regression models for physics grades regressed on collaborative variables, controlling for confounding variables.} 
  \label{models2020} 
\scriptsize 
\begin{tabular}{@{\extracolsep{5pt}}lccc} 
\\[-1.8ex]\hline 
\hline \\[-1.8ex] 
 & \multicolumn{3}{c}{\textit{Dependent variable:}} \\ 
\cline{2-4} 
\\[-1.8ex] & \multicolumn{3}{c}{Physics Grades 2020} \\ 
\\[-1.8ex] & (1) & (2) & (3)\\ 
\hline \\[-1.8ex] 
 Degree Collaboration & 0.199$^{**}$ &  &  \\ 
  & (0.092) &  &  \\ 
  & & & \\ 
 Weak Collaboration &  & 0.012 &  \\ 
  &  & (0.085) &  \\ 
  & & & \\ 
 Strong Collaboration &  & 0.224$^{**}$ &  \\ 
  &  & (0.100) &  \\ 
  & & & \\ 
 Degree Friendship &  &  & 0.048 \\ 
  &  &  & (0.109) \\ 
  & & & \\ 
 Weak Col. w/Prestige &  &  & 0.059 \\ 
  &  &  & (0.088) \\ 
  & & & \\ 
 Strong Col. w/Prestige &  &  & 0.145 \\ 
  &  &  & (0.125) \\ 
  & & & \\ 
 School 1 & $-$0.595$^{***}$ & $-$0.550$^{***}$ & $-$0.520$^{***}$ \\ 
  & (0.166) & (0.180) & (0.192) \\ 
  & & & \\ 
 Gender (Female) & 0.138 & 0.148 & 0.177 \\ 
  & (0.163) & (0.162) & (0.162) \\ 
  & & & \\ 
 Physics Grades 2019 & 0.357$^{***}$ & 0.343$^{***}$ & 0.352$^{***}$ \\ 
  & (0.091) & (0.092) & (0.094) \\ 
  & & & \\ 
 Constant & 0.247$^{*}$ & 0.216 & 0.190 \\ 
  & (0.142) & (0.148) & (0.151) \\ 
  & & & \\ 
\hline \\[-1.8ex] 
Observations & 99 & 99 & 99 \\ 
R$^{2}$ & 0.447 & 0.453 & 0.446 \\ 
Adjusted R$^{2}$ & 0.423 & 0.424 & 0.410 \\ 
Residual Std. Error & 0.767 (df = 94) & 0.767 (df = 93) & 0.776 (df = 92) \\ 
F Statistic & 18.990$^{***}$ (df = 4; 94) & 15.399$^{***}$ (df = 5; 93) & 12.335$^{***}$ (df = 6; 92) \\ 
\hline 
\hline \\[-1.8ex] 
\textit{Note:}  & \multicolumn{3}{r}{$^{*}$p$<$0.1; $^{**}$p$<$0.05; $^{***}$p$<$0.01} \\ 
\end{tabular} 
\end{table*}

\begin{table*}[!htbp] \centering 
  \caption{Difference and difference procedure.} 
  \label{did} 
\scriptsize 
\begin{tabular}{@{\extracolsep{5pt}}lcccccc} 
\\[-1.8ex]\hline 
\hline \\[-1.8ex] 
 & \multicolumn{6}{c}{\textit{Dependent variable:}} \\ 
\cline{2-7} 
\\[-1.8ex] & \multicolumn{6}{c}{Physics Grades} \\ 
\\[-1.8ex] & (1) & (2) & (3) & (4) & (5) & (6)\\ 
\hline \\[-1.8ex] 
 Gender (Female) & $-$0.032 & 0.046 & $-$0.032 & 0.071 & 0.008 & $-$0.029 \\ 
  & (0.154) & (0.148) & (0.148) & (0.148) & (0.155) & (0.145) \\ 
  & & & & & & \\ 
 Experimental Group & 0.212 & 0.147 & 0.335$^{**}$ & 0.141 & 0.139 & 0.349$^{**}$ \\ 
  & (0.149) & (0.149) & (0.154) & (0.148) & (0.156) & (0.153) \\ 
  & & & & & & \\ 
 Activity 3 & 0.317$^{**}$ & 0.353$^{**}$ & 0.282$^{*}$ & 0.350$^{**}$ & 0.305$^{**}$ & 0.297$^{**}$ \\ 
  & (0.150) & (0.150) & (0.148) & (0.148) & (0.152) & (0.146) \\ 
  & & & & & & \\ 
 Degree Collaboration & $-$0.257$^{***}$ &  &  &  &  &  \\ 
  & (0.095) &  &  &  &  &  \\ 
  & & & & & & \\ 
 Weak Col. &  &  & $-$0.409$^{***}$ &  &  & $-$0.438$^{***}$ \\ 
  &  &  & (0.118) &  &  & (0.118) \\ 
  & & & & & & \\ 
 Weak Col. w/Prestige &  &  &  &  & $-$0.049 &  \\ 
  &  &  &  &  & (0.108) &  \\ 
  & & & & & & \\ 
 Degree Friendship & 0.132 &  & 0.017 &  & 0.144$^{*}$ &  \\ 
  & (0.083) &  & (0.087) &  & (0.081) &  \\ 
  & & & & & & \\ 
 Physics Prestige & 0.056 &  &  &  &  &  \\ 
  & (0.078) &  &  &  &  &  \\ 
  & & & & & & \\ 
 Act.3*D. Collaboration & 0.277$^{*}$ &  &  &  &  &  \\ 
  & (0.152) &  &  &  &  &  \\ 
  & & & & & & \\ 
 Strong Col. &  & 0.020 &  &  &  & $-$0.049 \\ 
  &  & (0.101) &  &  &  & (0.082) \\ 
  & & & & & & \\ 
 Act.3*Strong Col.  &  & 0.099 &  &  &  &  \\ 
  &  & (0.152) &  &  &  &  \\ 
  & & & & & & \\ 
 Act.3*Weak Col.  &  &  & 0.297$^{**}$ &  &  & 0.291$^{**}$ \\ 
  &  &  & (0.146) &  &  & (0.146) \\ 
  & & & & & & \\ 
 Strong Col. w/Prestige &  &  &  & 0.074 &  &  \\ 
  &  &  &  & (0.101) &  &  \\ 
  & & & & & & \\ 
 Act.3*Strong Col. w/Prestige &  &  &  & 0.106 &  &  \\ 
  &  &  &  & (0.149) &  &  \\ 
  & & & & & & \\ 
 Act.3*Weak Col. w/Prestige &  &  &  &  & 0.166 &  \\ 
  &  &  &  &  & (0.150) &  \\ 
  & & & & & & \\ 
 Constant & $-$0.258$^{*}$ & $-$0.279$^{*}$ & $-$0.265$^{*}$ & $-$0.283$^{*}$ & $-$0.225 & $-$0.281$^{*}$ \\ 
  & (0.149) & (0.148) & (0.148) & (0.147) & (0.151) & (0.143) \\ 
  & & & & & & \\ 
\hline \\[-1.8ex] 
Observations & 180 & 180 & 180 & 180 & 180 & 180 \\ 
R$^{2}$ & 0.095 & 0.048 & 0.117 & 0.059 & 0.063 & 0.118 \\ 
Adjusted R$^{2}$ & 0.058 & 0.021 & 0.086 & 0.032 & 0.030 & 0.088 \\ 
Residual Std. Error & 0.970 (df = 172) & 0.990 (df = 174) & 0.956 (df = 173) & 0.984 (df = 174) & 0.985 (df = 173) & 0.955 (df = 173) \\ 
F Statistic & 2.586$^{**}$ (df = 7; 172) & 1.751 (df = 5; 174) & 3.813$^{***}$ (df = 6; 173) & 2.188$^{*}$ (df = 5; 174) & 1.924$^{*}$ (df = 6; 173) & 3.874$^{***}$ (df = 6; 173) \\ 
\hline 
\hline \\[-1.8ex] 
\textit{Note:}  & \multicolumn{6}{r}{$^{*}$p$<$0.1; $^{**}$p$<$0.05; $^{***}$p$<$0.01} \\ 
\end{tabular} 
\end{table*}





\end{document}